\documentclass[prl,aps,amssymb,twocolumn,showpacs,floatfix]{revtex4}

\usepackage{graphics}

\usepackage{epsfig} 

\usepackage{dcolumn}

\usepackage{amsmath}

\begin{document}

\title{Zero bias anomaly out of equilibrium}
\author{D. B. Gutman$^1$, Yuval Gefen$^2$, and A. D. Mirlin$^{3,4,*}$}
\affiliation{$^1$Dept. of Physics, University of Florida,
Gainesville, FL 32611, USA \\
$^2$Dept. of Condensed Matter Physics, Weizmann Institute of
  Science, Rehovot 76100, Israel\\
$^3$ Institut f\"ur Nanotechnologie, Forschungszentrum Karlsruhe,
 76021 Karlsruhe, Germany\\
$^4$ Inst. f\"ur Theorie der kondensierten Materie,
 Universit\"at Karlsruhe, 76128 Karlsruhe, Germany
}

\date{\today}

\begin{abstract}
The non-equilibrium zero bias anomaly (ZBA) in the tunneling density
of states of a 
diffusive metallic film
is studied. An effective action describing {\it virtual} fluctuations
out-of-equilibrium  is derived. 
The singular behavior of the equilibrium ZBA is
smoothed out by {\it real} processes of inelastic scattering.
\\[-0.5cm]
\end{abstract}

\pacs{73.23.-b, 73.40.Gk, 73.50.Td\\[-0.8cm] }


\maketitle The suppression  of tunneling current at low bias due
to  electron-electron interaction is known as the zero bias
anomaly (ZBA). The theory of ZBA for disordered metals  at thermal
equilibrium has been developed, on a perturbative level, by
Altshuler and Aronov \cite{AA,AA-review}. The non-perturbative
generalization of this theory was achieved by Finkelstein
\cite{Finkelstein83}. Measurements of the tunneling density of
states (DOS) in biased quasi-one-dimensional wires \cite{Pothier}
call for an extension of the theory to {\it non-equilibrium}
setups. In this work we study the ZBA for disordered metallic
films out of equilibrium, in both  the perturbative and the
non-perturbative (in interaction) limits.

Besides the experimental motivation, the problem  of ZBA
in a non-equilibrium system is of fundamental theoretical interest.
At equilibrium, the distribution of electrons in
phase space has a single edge at the Fermi surface.
The Coulomb interaction between the tunneling
electron and the electrons in the Fermi sea excites
virtual particle-hole pairs around the Fermi edge, leading to the
suppression of the tunneling DOS, similarly to the
Debye-Waller factor. The suppression gets stronger when the electron energy
approaches the Fermi energy.
Out of equilibrium, the distribution of particles may have
several sharp edges rather than a single one at the Fermi surface,
which poses important questions addressed in this work:
How will the excitation of electron-hole pairs in this situation
affect the tunneling DOS? Will there be an interpaly between the two edges?
We show that the two edges are not independent: one edge
affects the ZBA near the other  via real interaction-induced
scattering processes governing the dephasing of electrons in the
non-equlibrium regime. From this point of view the problem we are
considering is a representive of a class of phenomena  that
involve renormalization away from thermal equilibrium,
such as the Fermi edge singularity \cite{Abanin} and
the Kondo effect \cite{Paaske}.

What makes the ZBA particularly interesting is its deep connection to
various conceptually important phenomenological ideas.
At {\it equilibrium}, the non-perturbative results \cite{Finkelstein83}
have been reproduced by  quantum hydrodynamical
methods \cite{Levitov}, and, within the framework of the theory of
dissipation \cite{Nazarov89}, by
methods that rely on the fluctuation-dissipation theorem.
Our work circumvents this restriction.
Starting with the Keldysh non-linear $\sigma$-model, we derive an effective
action that
accounts for virtual fluctuations  in disordered metals away from
equilibrium.  This  action  is complementary to the
one for  kinetics of real fluctuations
(such as noise)   developed earlier
\cite{GMG,bodineau04,pilgram04a,Bagrets}.
Further, we discuss a connection between our theory and
phenomenological methods \cite{Levitov,Nazarov89,Devoret91,Glazman91}.
As a central application of our theory, we analyze the ZBA problem
and calculate the tunneling DOS
for a two-dimensional (2D) diffusive metallic film subject to external bias.

Consider an  electron which  tunnels from a tip into a metal,
subject to an external bias. The Coulomb  interaction $U(q)$
($U(q)= 2\pi e^2/q$, $d=2$) causes  electrons in the metal to
readjust their position, such that they try  to screen the added
charge. Their motion  is impeded  by  static disorder. For a
sufficiently good metal, characterized by the dimensionless
conductance $g=\epsilon_F \tau \gg 1$, ($\epsilon_F$ is  the Fermi
energy and $\tau$ -- the elastic mean free time) the electron
motion is diffusive. 
Low energy excitations are accounted for by
the non-linear Keldysh $\sigma$-model \cite{Kamenev-Andreev}
\begin{eqnarray}
\hspace*{-1cm}&&
iS[Q,\phi]=i{\rm Tr}\{\phi^T(U^{-1}+\nu_0)\sigma_1\phi\} \nonumber \\
\hspace*{-1cm} &&
-\frac{\pi\nu_0}{4}{\rm Tr}\{D(\nabla Q)^2-4\partial_t Q\}
-i\pi\nu_0{\rm Tr}\{\phi_\alpha\gamma^\alpha Q\},
\end{eqnarray}
where $Q({\bf r},t,t')$ is a $2\times 2$ matrix field (in the
Keldysh space)  describing slow electronic modes, $\phi({\bf
r},t)$ is a 2-vector in the Keldysh space  representing the
Coulomb interaction, and ${\rm Tr}$ denotes the trace both over
Keldysh indices (${\rm Tr}_K$), spatial, and temporal coordinates.
The $Q$ field is subject to a non-linear constraint,
\begin{eqnarray}
\int dt_1 Q({\bf r},t,t_1)Q({\bf r},t_1,t')=\delta(t-t')\sigma_0,
\end{eqnarray}
$\gamma_1\equiv\sigma_0$ is a $2\times 2$ unit matrix,
$\gamma_2\equiv\sigma_1$ the 
first Pauli matrix,
$\nu_0$ is the DOS at the Fermi energy in the absence of interaction,
and $D$ is the diffusion coefficient.
The theory can be rewritten in terms of the field $Q$ only.
Integrating the $\phi$ field  out, we derive the following action:
\begin{eqnarray}&&
\label{e1}
iS[Q]=-\frac{\pi\nu_0}{4}\left({\rm Tr}\{D(\nabla
  Q)^2-4\partial_tQ\}\right) \nonumber \\&&
-\frac{i(\pi\nu_0)^2}{2}{\rm Tr}_K\{\sigma_1
Q\}\left(U^{-1}+\nu_0\right)^{-1}{\rm Tr}_K\{Q\}.
\end{eqnarray}
Physical observables can be found by  differentiating the generating functional
\begin{eqnarray}
\hspace*{-1cm} &&
Z[\varphi_1,\varphi_2]=\exp\left(iS[Q]\right)
\exp\left(\frac{i\nu_0}{1+\nu_0 U} \right.
\nonumber
\\ \hspace*{-1cm} &&  \left.
\times \bigg[2\varphi_1\varphi_2-\pi\varphi_1{\rm Tr}_K\{Q\}
-\pi\varphi_2{\rm Tr}_K\{\sigma_1Q\}\bigg]\right)
\end{eqnarray}
over the source fields $\varphi_1$ and $\varphi_2$.

So far all low energy excitation in the problem
have been kept indiscriminately. The  price one pays for this is an
extreme technical complexity of the theory.
In some cases, this description is excessive  and the theory can be
simplified  by singling out a particular subspace of the  $Q$ matrices.
The resulting  theory is less general,
but more suitable  for tackling a specific class of problems.
The noise statistics in disordered systems is a remarkable  example of such
a simplification.
An extension of the Boltzman-Langevin approach \cite{KTF}
to  high order correlators
is achieved  on the subspace of matrices that
are diagonal in the Wigner representation,
\begin{eqnarray}
\label{ansatz-real}
Q({\bf r},\epsilon,t)=
e^{\frac{i\sigma_1}{4}\bar{f}({\bf r},\epsilon,t)}
\left(\!\!\begin{array}{cc}
1&  2-4f({\bf r},\epsilon,t) \\
0& -1
\end{array}\!\!\right)
e^{-\frac{i\sigma_1}{4}\bar{f}({\bf r},\epsilon, t)}. \nonumber \\
\end{eqnarray}
The theory \cite{pilgram04a,GMG,Bagrets} resulting  from the
$\sigma$-model (\ref{e1}), reduced to the subspace
(\ref{ansatz-real}), accounts for real fluctuations in agreement
with the cascade idea of Nagaev \cite{Nagaev}.

Below  we use a similar ideology, ``projecting'' the $\sigma$-model on
the subspace appropriate for the ZBA problem.
According to works on ZBA at equilibrium, these are
virtual fluctuations of gauge fields that
dominate the suppression of the tunneling DOS
\cite{Finkelstein83,Kopietz98,Kamenev-Andreev}.
This suggests that
gauge-type fluctuations constituting local-in-time
rotations of the saddle point $\Lambda_{t-t'}$,
\begin{eqnarray}
\label{ansatz}
\hspace*{-0.5cm} && Q({\bf r},t,t') = e^{\frac{i\sigma_1}{2}\bar{f}({\bf
    r},t)+\frac{i}{2}f({\bf r},t)}\Lambda_{{\bf r},t-t'}
e^{-\frac{i}{2}f({\bf r},t')-\frac{i\sigma_1}{2}\bar{f}({\bf
    r},t')}, \nonumber \\
\hspace*{-0.5cm} && \Lambda_{{\bf r},t-t'} = \left(\begin{array}{cc}
1& 2-4n({\bf r},t-t') \\
0& -1
\end{array}\right)\:,
\end{eqnarray}
where $n({\bf r},t-t')$ is the Fourier transform of the local distribution
function $n({\bf r},\epsilon)$,
are to be retained. Plugging (\ref{ansatz}) into  the action
(\ref{e1}) and expanding up to quadratic order in $f$ and $\bar{f}$,
we derive  an {\it effective theory of fluctuations} in an
interacting diffusive conductor.
\begin{eqnarray}
\label{e3}
 && iS[f,\bar{f}]=\frac{\nu_0}{2}\int (d\omega)(dq)\nonumber \\
&& \times \bigg[\bar{f}(-\omega,-q)\:\omega \left(-Dq^2+\frac{i\omega}{1+\nu_0
    U(q)}\right)f(\omega,q) \nonumber \\
&& -Dq^2T(\omega)\bar{f}(-\omega,-q)\bar{f}(\omega,q)
\bigg]\, .
\end{eqnarray}
Here the effective temperature  $T(\omega)$  is defined as
\begin{equation}
\label{e7}
T(\omega)=\int_{-\infty}^{\infty} d\epsilon\:
n(\epsilon)\:[2-n(\epsilon-\omega)-n(\epsilon+\omega)]\,,
\end{equation}
and $n(\epsilon)$ is a distribution function determined by the
Boltzmann equation.

To demonstrate the consistency of our approach, we first show  that
the effective action (\ref{e3}) reproduces correctly known density
correlation functions. The symmetrized density correlation function is given by
\begin{eqnarray}
\hspace*{-0.7cm}&&
{1\over 2} \langle \big[\hat{\rho}({\bf
  r},t),\hat{\rho}(0,0)\big]_+\rangle_{\omega,q}=
-\frac{1}{4}\frac{\partial^2 Z}
{\partial \phi_2(\omega,q)\partial\phi_2(-\omega,-q)} \nonumber \\
\hspace*{-0.7cm} &&
=\frac{1}{4}\left(\frac{\nu_0}{1+\nu_0
    U}\right)^2\omega^2\langle f f \rangle_{\omega,q} \, .
\label{e101}
\end{eqnarray}
Similarly,  the  response function can be expressed as
\begin{eqnarray}
\hspace*{-0.7cm} &&
i\langle [\hat{\rho}({\bf r},t),\hat{\rho}(0,0)]_-\theta(t)\rangle_{\omega,q}=
\frac{1}{2i}\frac{\partial^2 Z}{\partial
  \phi_1(\omega,q)\partial\phi_2(-\omega,-q)}
\nonumber \\ \hspace*{-0.7cm} &&
=\frac{i}{2}\left(\frac{\nu_0}{1+\nu_0 U}\right)^2 \omega^2
\langle f \bar{f}\rangle_{\omega,q}+\frac{\nu_0}{1+\nu_0 U}\,.
\label{e102}
\end{eqnarray}
Calculating  the correlation functions entering Eqs.~(\ref{e101}),
(\ref{e102}) using the action (\ref{e3}),
\begin{eqnarray}&&
\langle f f \rangle_{\omega,q}=
\frac{4Dq^2T(\omega)}{\nu_0
  \omega^2\left|Dq^2+\frac{i\omega}{1+\nu_0
      V(q)}\right|^2}\:, \nonumber\\&&
\langle f_{\omega,q} \bar{f}_{-\omega,-q}\rangle \equiv
\langle f \bar{f}\rangle_{\omega,q}=
\frac{2}{\nu_0}
\frac{1}{\omega\left(Dq^2 - \frac{i\omega}{1+\nu_0 U(q)}\right)},
\label{e103}
\end{eqnarray}
we reproduce the known results for
for  the spectral function of density fluctuations
\begin{eqnarray}&&
{1\over 2}\langle \big[\hat{\rho}({\bf
  r},t),\hat{\rho}(0,0)\big]_+\rangle_{\omega,q}=\frac{\nu_0 Dq^2
  T(\omega)}{|Dq^2[1+\nu_0 U(q)]+i\omega|^2}\,,\nonumber
\end{eqnarray} and 
the density-density response function,
\begin{eqnarray}&&
i\langle [\hat{\rho}({\bf r},t),\hat{\rho}(0,0)]_-
\theta(t)\rangle_{\omega,q}=
\frac{\nu_0 Dq^2}{Dq^2[1+\nu_0 U(q)]-i\omega}\, .\nonumber
\end{eqnarray}
It is worth noting the analogy between  our effective action and
phenomenological theories that describe ZBA  at equilibrium within
an effective environment model \cite{Glazman91,Devoret91}.
The latter explain the ZBA as the influence of
virtual fluctuations of an electromagnetic field at equilibrium
\cite{Nazarov89} on the electron tunneling.
In view of the fluctuation-dissipation theorem,
fluctuations of the  electromagnetic field
are determined by the complex impedance of the system.
In the zero dimensional case,  modes of the electromagnetic field
can be considered  as independent quantum harmonic oscillators.
Being suddenly shaken by the incoming electron, they move away
from the equilibrium  position, reducing the overlap with their
original configuration hence suppressing electron tunneling
amplitude.
The action (\ref{e3}) describes similar processes in a 2D diffusive
system and without the assumption of  thermal equilibrium, thus
keeping the information about both real and virtual processes in a
non-equilibrium state.

Now we are ready to apply the theory to the problem of ZBA out of equilibrium.
The tunneling DOS
\begin{eqnarray}
\nu(\epsilon)=\frac{i}{2\pi}\int
(dp)\left(G^R(\epsilon,p)-G^A(\epsilon,p)\right)
\end{eqnarray}
can be rewritten in terms of  $Q$ matrices as
\begin{eqnarray}
\label{e2}
\nu(\epsilon)=\frac{\nu}{2}\int_{-\infty}^\infty \!\! dt \:
e^{-i\epsilon(t-t')}\langle {\rm
  Tr}_K\ Q({\bf r},t',t)\sigma_z \rangle_Q\,.
\end{eqnarray}
Plugging Eq.~(\ref{ansatz}) into Eq.(\ref{e2}), we find
\begin{eqnarray}&&
\label{e5}
\hspace{-1cm}\frac{\nu(\epsilon)}{\nu_0}=1\!+\!2i\int_{-\infty}^{\infty}
\!\!dt \,n(t)\:
e^{i\epsilon t-\frac{1}{4}I_{ff}(t)}
\sin\left(\frac{1}{2}I_{f\bar{f}}(t)\right),
\end{eqnarray}
where we used the notations
\begin{eqnarray}&&
\label{d1}
I_{ff}(t)=\int(d\omega)(dq)\:(1-\cos \omega t)\:\langle\langle f
f\rangle\rangle_{\omega,q}\ , \nonumber  \\&&
I_{f\bar{f}}(t)=\int (d\omega)(dq)\:\sin \omega t\;\langle\langle f
\bar{f}\rangle\rangle_{\omega,q}\, .
\end{eqnarray}
There is a subtle point related to the definition of the
correlator $\langle\langle ff\rangle\rangle$ and $\langle\langle
f\bar{f}\rangle\rangle$ in (\ref{d1}). The ZBA should vanish in
the absence of interaction ($\nolinebreak{U=0}$). This physically
obvious statement is valid to all orders of the diagrammatic
calculations (all potentially singular contributions vanish after
the time integration over the Keldysh contour) and is satisfied by
the non-linear $\sigma$ model. However, it ceases to be an exact
feature of the theory when the full space of $Q$ matrices is
reduced to the subspace Eq.~(\ref{ansatz}). This problem is easily
cured. Since we are only interested in summing up the leading
interaction-induced $\ln^2 \epsilon$ contributions to the
tunneling DOS (to all orders), these are the interaction-induced
parts of $\langle ff\rangle$ and $\langle f\bar{f}\rangle$ that we
have to keep in Eq.~(\ref{d1}) \cite{subtraction}:
\begin{eqnarray}
\label{e4a}
\hspace*{-0.7cm}
\langle \langle f\bar{f}\rangle\rangle_{\omega,q}
& \equiv & \langle f\bar{f}\rangle_{\omega,q}
-\langle f\bar{f}\rangle_{\omega,q,U=0} \nonumber \\
& = &
\frac{-2iU}{(Dq^2-i\omega)[Dq^2(1+\nu_0 U)-i\omega]}\,.
\end{eqnarray}
Analogously,
\begin{eqnarray}
\hspace*{-1cm} &&
\label{e4}
\langle\langle ff\rangle\rangle_{\omega,q}=
\frac{4Dq^2 T(\omega)(2+\nu_0 U)U}{|Dq^2-i\omega|^2|Dq^2(1+\nu_0
  U)-i\omega|^2}\,.
\end{eqnarray}
Equations (\ref{e5})--(\ref{e4}) constitute our general result for the
non-equilibrium ZBA. At equilibrium they reproduce the known results for
the ZBA in the perturbative \cite{AA,AA-review} and non-perturbative
\cite{Finkelstein83} regimes.

To  illustrate the effect of non-equilibrium conditions on ZBA we
consider a diffusive film of length $L$
connected to two zero-temperature reservoirs  with voltage difference
$V$.
We assume that the energy relaxation in the film can be neglected,
i.e. that $L$ is shorter than the energy relaxation length, a condition  met
at not too high bias.
In this case the solution of the Boltzmann equation is a double-step function
\begin{equation}
n(\epsilon,x)=a
n_0(\epsilon-eV/2)+(1-a)n_0(\epsilon+eV/2)
\,,
\label{double-step}
\end{equation}
where $n_0(\epsilon)$ is the  Fermi distribution function,
$a=x/L$, and $x$ is the distance from the reservoir.
This distribution  corresponds to the effective temperature
\begin{eqnarray}&&
T(\omega)= [a^2+(1-a)^2 ] T_{\rm eq}(\omega) \nonumber \\&&
+ a(1-a) [T_{\rm eq}(\omega+eV)+T_{\rm eq}(\omega-eV)] \,,
\end{eqnarray}
where $T_{\rm eq}(\omega)$ is the equilibrium value
\begin{eqnarray}
T_{\rm eq}(\omega)=\omega\coth (\omega/T)   \mathop {\rightarrow
}\limits_{T \to 0 }|\omega|\,.
\end{eqnarray}
Substituting Eqs.~(\ref{e4a}), (\ref{e4}) into Eqs.(\ref{d1}),(
\ref{e5}), we find the tunneling DOS  out of equilibrium:
\begin{eqnarray}
\label{e6}
\hspace*{-0.8cm} &&  \frac{\nu(\epsilon)}{\nu_0}=a\exp\left(
-\frac{1}{8\pi^2\nu_0 D}
\log{t_-\over \tau}\log(D^2\kappa^4\tau t_-)\right) \nonumber\\
 \hspace*{-0.8cm} &&
+ (1-a)\exp\left(-\frac{1}{8\pi^2\nu_0
    D}\log{t_+\over \tau}
\log(D^2\kappa^4\tau t_+)\right)\,.
\end{eqnarray}
This result is valid in both the perturbative (when the exponentials
can be expanded up to the first non-trivial term) and non-perturbative
regimes.
The DOS consists of two terms  corresponding to electrons coming  from
the left and right reservoir.
The energy scales governing the argument of the logarithm in these two
terms are
\begin{eqnarray}
\label{e7}
t_\pm^{-1}={\rm
  max}\bigg\{\left|\epsilon\pm\frac{eV}{2}\right| ,
\frac{a(1-a) eV\log\left(\frac{\nu_0 D^2\kappa^2}{eV}\right)}
{4\pi\nu_0 D}
\bigg\},
\end{eqnarray}
respectively ($\kappa=2\pi e^2\nu$).

\begin{figure}
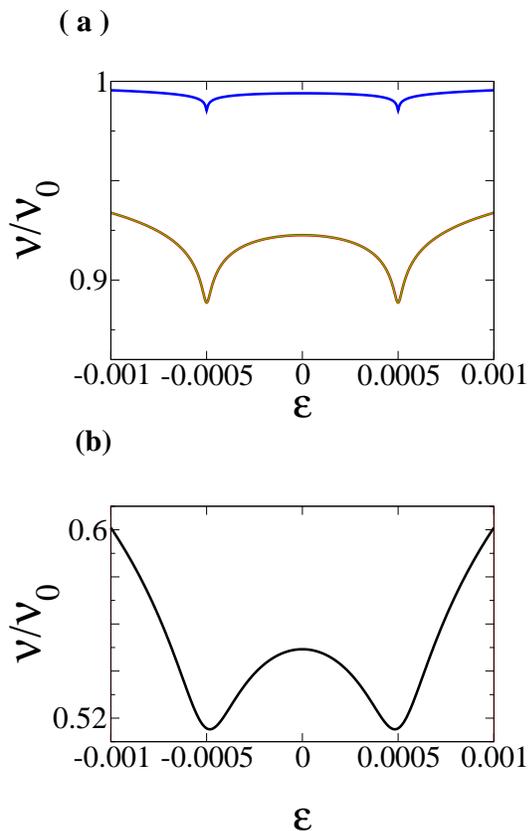

\includegraphics[width =0.8\columnwidth]{inter_and_pert1.eps}

\vskip0.1cm

\includegraphics[width = 0.8\columnwidth]{zba-nonpert1.eps}

\caption{Tunneling DOS vs.  energy for the
double-step distribution function (\ref{double-step}) with $a=0.5$.
The dimensionless conductance is, from top to bottom,
$g=100$, 10 (a) and 1 (b).}
\label{fig1}
\vspace*{-0.5cm}
\end{figure}

The evolution of the ZBA in the tunneling DOS with decreasing
conductance (i.e. from the perturbative to the non-perturbative
regime) is illustrated in Fig.~\ref{fig1}. The DOS has a two-dip
shape with  minima reached at the energies where the distribution
function exhibits discontinuous jumps (i.e. at the positions of
the Fermi edge  in the  left and right leads, $\epsilon =
\pm\frac{eV}{2}$). Away from the minima, the DOS is controlled by
the energy measured from the corresponding edge. As  this energy
decreases (we get closer to one of the Fermi edges), the
singularity near this edge gets affected by the presence of the
other edge. The broadening of the ZBA singularity takes place on a
new energy scale,
\begin{eqnarray}
\label{tauphi}
\tau^{-1}_\phi (V) = \frac{a(1-a)}{4\pi\nu_0 D}
eV\log\left(\frac{\nu_0D^2\kappa^2}{eV}\right)\,.
\end{eqnarray}
The notation introduced in Eq.~(\ref{tauphi}) stresses
the analogy with the equilibrium
dephasing rate $\tau_\phi^{-1} (T)$  governing the
infrared cutoff of the interference phenomena such as the weak
localization \cite{AA-review}.
The emergence of $\tau^{-1}_\phi (V)$ shows that inelastic scattering
processes (responsible for dephasing) lead to smearing of the ZBA
singularity.

Qualitatively the results can be explained as follows.
Virtual fluctuations of the gauge fields act on a quasiparticle
and suppress  the tunneling DOS.
Their influence is limited  by the  quasiparticle life time.
An external  bias  enhances  fluctuations of the electromagnetic
field, which in turn dephase the tunneling quasiparticle. This
results in a competition between the virtual processes
establishing the ZBA and the real fluctuations (noise) cutting it
off.

Let us stress the crucial difference with the equilibrium
situation. At equilibrium, the ZBA singularity in 2D is cut off by
temperature (which enters via the distribution function);
specifically,  the dephasing rate $\tau_\phi^{-1}(T)$ does not
affect the ZBA in any essential way, since $\tau_\phi^{-1}(T)\ll
T$. It is thus a distinct feature of the strongly non-equilibrium
regime that the ZBA is smeared by the inelastic scattering
(dephasing) rate.

In conclusion, we have developed an  effective theory for virtual
fluctuations in disordered metals out of equilibrium.  Using this
theory, we studied the ZBA in a 2D metallic film biased by
external voltage. We have found that out of equilibrium the
tunneling DOS has a double-dip structure with minima reached at
the "edges" of the particle distribution. The ZBA near any of the
"edges" is influenced by the other one. The suppression of DOS is
smoothed out by the real processes of inelastic scattering
(dephasing) with characteristic energy scale $\tau_{\rm \phi}(V)$.
Further applications and extensions of our theory include, in
particular, the ZBA in quasi-one-dimensional and strictly
one-dimensional (Luttinger liquid) wires \cite{Mishchenko} out of
equilibrium; these results will be reported elsewhere.

We thank D. Bagrets, N. Birge, A. Finkelstein, D. Maslov,  A.
Shnirman, and A. Shytov for useful discussions. This work was supported by
NSF-DMR-0308377 (DG), US-Israel BSF, ISF of the Israel
Academy of Sciences, and DFG SPP 1285 (YG), DFG Center for
Functional Nanostructures, EC Transnational Access Program at the
WIS Braun Submicron Center (ADM), and Einstein Minerva Center.

\end{document}